\documentclass[12pt]{article}
\usepackage[utf8]{inputenc}
\usepackage{textcomp}
\usepackage{newunicodechar}
\newunicodechar{−}{-}
\usepackage{amsmath}
\usepackage{amsfonts}
\usepackage{graphicx}
\usepackage{orcidlink}
\usepackage{xcolor}
\usepackage[normalem]{ulem}
\graphicspath{ {images/} }
\usepackage[margin=1in,top=1.5in,bottom=1.5in]{geometry}

\title{Volatility Forecasts with Market Trends}
\author{
  Sara A. Safari$^{1,2}$\,\orcidlink{0000-0003-0376-7429} 
  \and Christof Schmidhuber$^{1}$\,\orcidlink{0000-0001-6068-4027}
}

\begin{document}

\maketitle
\footnotetext[1]{\small Zurich University of Applied Sciences, School of Engineering, 
  Technikumstrasse 9, CH-8401 Winterthur, Switzerland}
\footnotetext[2]{\small University of Zurich, Department of Mathematical Modeling 
  and Machine Learning ($DM^3L$), Winterthurerstrasse 190, CH-8057 Zurich, Switzerland}

\thispagestyle{empty}

\begin{abstract}
We show how volatility forecasts for financial markets can be refined by 
accounting for the current trends in these markets. 
Empirically, volatilities tend to increase day after day in times of strong 
up- or down-trends. This effect is particularly pronounced during down-trends. 
We find that it can accurately be modeled by including quadratic polynomials of the current trend strength as explanatory factors. 
Such factors can improve the predictive power
of a wide range of volatility models, ranging from linear EWMA-based models over GARCH models and range-based models to stochastic volatility models. 
Our results are based on a robust statistical analysis of three decades of daily futures prices
across equity-, bond-, FX- and commodity markets.

\end{abstract}

\thispagestyle{empty}

\newpage
\section{Introduction}
\label{sect:ref}

Over the past decades, market trends have proven to have a tiny but still statistically significant predictive power for future returns.
This effect has been very profitably exploited by the trend-following industry at least since the turtle traders in the 1980's. It has also been confirmed in many academic studies, such as 
\cite{cutler, erb, miff, mosk, baz}. More recently, it has been shown that the predictive power of market trends is not a new phenomenon but dates back at least a few centuries \cite{lemp, hurst, grey, sara}. \\

In \cite{lemp}, it has been noted that trends tend to revert once they become too strong, and this has been modeled quantitatively in \cite{black}.
In \cite{schmidhuber}, these results were independently confirmed by carefully measuring the next-day mean return in a market as a function of its trend strength on a given day.
To filter out the statistical noise and get sufficiently precise results, it was crucial
to aggregate not only across decades of daily market data for a broadly diversified set of 
financial markets, but also across 10 different trend horizons ranging from days to years. 
This aggregation makes sense, because - as is well-known in practice - similar trend-following strategies tend to
work over a wide range of time scales.\\

\begin{figure}[t]\centering
    \includegraphics[height=4.7cm]{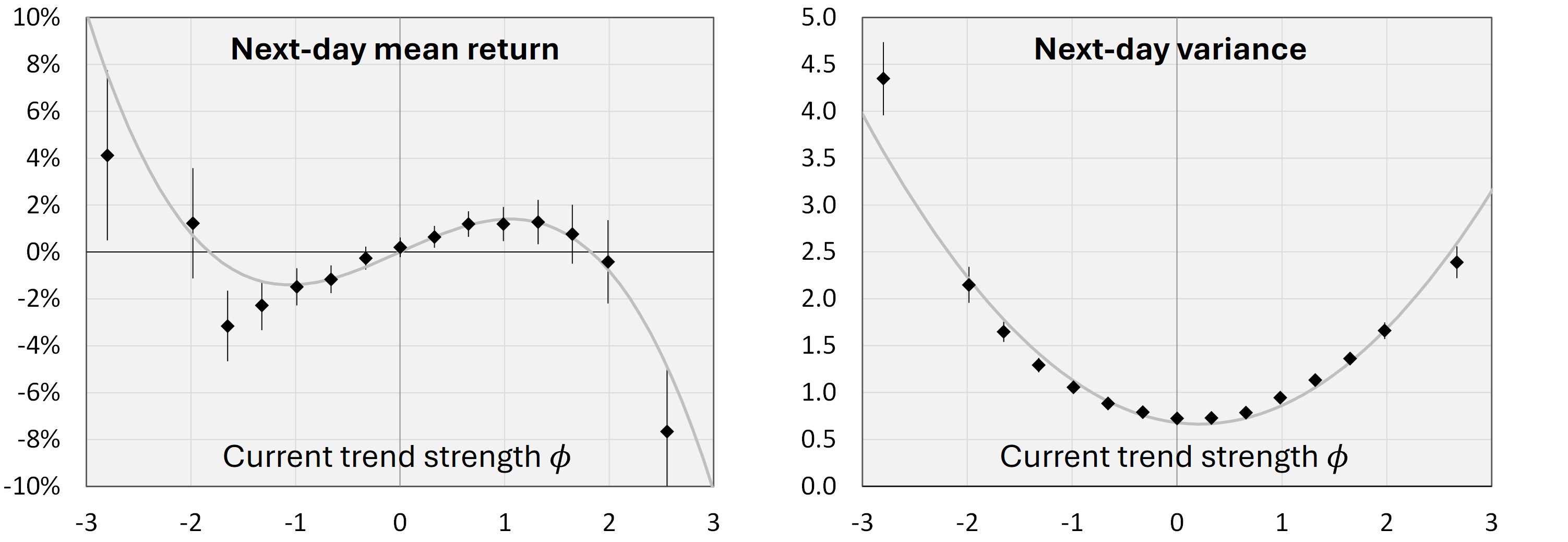}
    \caption{{\it Left}: The expectation value $E(r)$ of tomorrow's return in a futures market can be modeled by a cubic polynomial of the $t$-statistics $\phi$ of today's trend. 
The linear term models trend-persistence. The cubic term models trend-reversion.  
The large standard errors, indicated by bars, reflect the huge amount of noise in market returns.
{\it Right}: The variance of the next-day return is higher in times of strong trends and can be modeled by a parabola. 
}\label{figA}
\end{figure}

A key finding of \cite{schmidhuber} is illustrated in fig. 1 (left): 
trends tend to revert before they become statistically strongly significant. This means that as soon
as a trend has become so obvious that it is clearly visible in a price chart, it is already too late to join it. 
This is consistent with the hypothesis \cite{fama} that any obvious market inefficiency is quickly eliminated by investors. More precisely, for a given market and a given time horizon, 
one defines the strength $\phi$ of a trend as its $t$-statistics. 
It was found that tomorrow's return $r_{t+1}$ (normalized to have variance 1) is modeled
well by a cubic polynomial of today's trend strength $\phi_t$:
\begin{eqnarray}
r_{t+1}&=&a\ +{b}\cdot \phi_t+{c} \cdot \phi_t^3+\epsilon_{t+1},\label{cubic}\\
\text{with}&&b\ =\ 0.013\pm0.004\ ,\ \ \ c\ =\ -0.006\pm0.002,\notag
\end{eqnarray}
where the random noise $\epsilon$ is not required to be identically and independently distributed. 
Here, $a$ is a risk premium, which depends on the asset.
However, the kinetic coefficients $b$ and $c$ are universal in the sense that they seem to be 
the same for all assets within the limits of statistical significance.
$b$ is interpreted as the persistence of trends, and $c$, which is negative, as the strength of trend reversion.
Both are statistically highly significant, while quadratic and higher-order polynomial terms in (\ref{cubic}) are not or only weakly significant.  
Model (\ref{cubic}) is in line with practical experience of the trend-following industry.
It has been extended to intraday- and many-year-time scales \cite{lemp,sara}.\\

In this paper, we empirically analyze what is denoted as ``noise" $\epsilon$ 
in (\ref{cubic}) in more detail and find that it also has a rich and interesting structure. 
As a simple illustration, fig. 1 (right) plots the expectation value of tomorrow's square return 
$r_{t+1}^2$ against today's trend strength. This yields a slightly skewed parabola with only
small estimation errors.
It shows that the variance of tomorrow's return is much higher in times of strong up- or down-trends. \\

We will quantify and refine this observation
by precisely measuring the next day's expected variance as a function of both the trend strength 
and the variance on a given day. 
Our analysis is based on more than 35 years of daily futures returns for 24 different markets,
diversified across different asset classes (equity indices, interest rates, FX rates and commodities) 
and across different time zones (Americas, Europe and Asia). 
We use futures returns instead of index returns, as they are guaranteed to be marked-to-market daily.
We find that tomorrow's variance can be modeled well by 
today's variance, supplemented by
second-order polynomials in today's trend strengths. 
Including the latter as explanatory factors turns out to yield a moderate, 
but statistically highly significant increase of the predictive power
of simple mean-reverting volatility models as well as more advanced ones. \\

This paper is organized as follows. Section 2 describes our data, our trend definition, 
and our statistical methodology including bootstrapping and out-of-sample cross validation,
which follows that of \cite{schmidhuber}.
Section 3 presents our empirical measurements and analyses of the next-day variance
for a simple linear model based on EWMA volatility \cite{riskmetrics}.
We observe that taking current market trends into account
significantly improves the predictive power of the model, 
as measured by the out-of-sample $R$-squared in the case of
ordinary least squares (OLS) regression, and by the out-of-sample quasi-likelihood in the case of maximum likelihood (MLE) estimation. We also analyze how our results depend on 
the trend horizon and the asset class, assess their stability in time, and discuss longer-term predictions.\\

In section 4, we consider advanced volatility models, and we 
show that adding trends as explanatory factors can improve their predictive power as well.
Specifically, we consider the GARCH(1,1) model \cite{garch}, 
its GJR generalization \cite{gjr},
a GARCH model based on daily price ranges \cite{parkinson}, 
the HAR model \cite{corsi},
the Heston model \cite{heston}, and the leverage stochastic volatility (SV) model \cite{jacquier, yu}.
We find a moderate, but statistically highly significant improvement 
of the out-of-sample predictive power for all of these models. \\ 

Section 5 summarizes our results. They have the potential to help improve
market risk management tools by taking market trends into account, especially in periods of 
market crises, which are typically accompanied by strong trends. 
Beyond immediate practical applications, our empirical observations may also help in the search 
of a financial market model
that can derive the stylized facts of finance \cite{mant, cont}
from first principles.

\section{Data and Methodology}

\subsection{Markets and Data}

Our analysis is based on historical daily log-returns for the set of futures  markets shown in table 1. This set is diversified across four asset classes (equity indices, interest rates, currencies, commodities), 
across three regions (Americas, Europe, Asia), and across three commodity sectors (energy, metals, agriculture).  

\begin{table}[h!]
\centering
\begin{tabular}{ |p{3.3cm}||p{3.3cm}|p{3.3cm}|p{3.3cm}|  }	\hline
	{\bf Regions:}& {\bf America}  &{\bf Europe} &{\bf Asia}\\	\hline
	Equities   & S\&P 500    &DAX 30&  Nikkei 225\\
	&  TSE 60  & FTSE 100   & Hang Seng\\ \hline
	Interest rates &US 10-year & Germany 10-year &  Japan 10-year \\
	&Canada 10-year  & UK 10-year &  Australia 3-year \\	\hline
	Currencies&   CAD/USD  &  EUR/USD& JPY/USD\\
	&    & GBP/USD   &AUD/USD\\	& & &NZD/USD\\\hline\hline
	Commodities& Crude Oil  &Gold &Soybeans\\
	& Natural Gas  & Copper &Live Cattle\\	\hline
	{\bf Com.-Sectors:}  & {\bf Energy}  &{\bf Metals}   &{\bf Agriculture}\\	\hline
\end{tabular} 
\caption{The diversified set of futures markets used for our analyses.}
\end{table}

For all contracts, our data set consists of daily prices from Jan 1, 1990, to Sep 15, 2026. Daily prices $P_i(t)$ were taken from Bloomberg, where $i$ labels the asset. Futures are rolled 5 days prior to first notice. Our data are a slightly updated version of the data set \cite{mendel} used in \cite{schmidhuber} (the updated version is available upon request). 
The actual analysis begins in May 1991, covering $N=9216=9\cdot2^{10}$ daily log returns for each asset (for our analysis, it is convenient that $N$ is divisible by $2^{10})$. 
Earlier data are only used to compute the initial trend strengths and variances in May 1991.
Some early data had to be reconstructed as follows:
\begin{itemize}
\item The history of TSE 60 futures prices begins in September 1999. Before, they
were reconstructed from TSE 300 index prices, dividend yields, and Canadian Libor rates.
\item The history of Hang Seng index future prices only begins in April 1992. In the first year, they are 
proxied by Nikkei 225 futures. 
\item The history of EUR futures prices begins in May 1998. Before, they were
reconstructed from the Deutsche Mark/USD spot exchange rate and German/U.S. Libor differentials.
\item The history of NZD futures prices begins in May 1997. Before, they were reconstructed from the spot exchange rates and NZD/USD Libor differentials.
\end{itemize}
For the analysis of range-based models in subsection 4.3, we use daily High/Low prices for the same set of assets,
although those cover only $8192$ days since April 1995.

\subsection{Definition of Trend Strength and Variance}

We consider trends of 8 different time horizons:
\begin{equation}
T_k=2^k\ \text{business days with}\  k\in \{1,2,3,...,8\}\notag
\end{equation}
This represents periods of approximately 2 days, 4 days, 8 days, 3 weeks, 6 weeks, 3 months, 6 months, and 1 year. Thus, there are 8 different trend strengths at each point in time. A given asset may well be, e.g., in a long-term up-trend at the 1-year time scale, and at the same time in a short-term down-trend at the 3-week time scale. \\

We define the trend strengths as follows: 
Let $P_i(t)$ be the daily closing price of market $i$. We define normalized log-returns $R_i(t)$: 
\begin{equation}
R_i(t)={r_i(t)\over \sigma_i}\ ,\ \ \ r_i(t)=\ln {P_i(t)\over P_i(t-1)}\ ,\ \ \ \sigma_i^2=\text{var}(r_i)\ ,\ \ \ \mu_i=\text{mean}(r_i)\ ,\notag
\end{equation}
where $\mu_i$ is the long-term daily risk premium of a market $i$, and $\sigma_i$ is its long-term daily standard deviation. For a given horizon $T$ and a given market $i$, the 
trend strength $\phi_{i,T}(t)$ at the close of trading on day $t\in Z$ is a weighted 
average of previous log-returns of that market in excess of the long-term risk premium:
\begin{equation}
\phi_{i,T}(t)=\sum_{n=0}^\infty w_T(n)\cdot \hat R_i(t-n)\ \ \ \text{with}\ \ \ \hat R_i(t-n)=R_i(t-n)-{\mu_i\over\sigma_i}.\notag
\end{equation}
Here, $w_T(n)$ is a weight function (see \cite{levine} for different variants). We normalize $w_T$ such that the trend strength 
$\phi_{i,T}$ has standard deviation 1. Assuming that returns on different days are independent 
of each other (which is true to high accuracy), this implies: 
\begin{equation}
\sum_{n=0}^\infty w_T^2(n)=1.\notag
\end{equation}
With this normalization, $\phi_{i,T}$ is the t-statistics of the trend. 
All trend strengths are then comparable with each other, so we can aggregate them across all markets and time scales.
As in \cite{schmidhuber}, we use the following weight function for the trend strength $\phi_T$: 
\begin{eqnarray}
\tilde w_T(n)&=&M_T\cdot\ n\cdot e^{-2n/T} \ \ \ \text{with normalization factor} \ \ M_T.
\notag\end{eqnarray}
This ``smooth weight function" is compared in fig. 2 (left) with an equally-weighted and
an exponentially weighted moving average (EWMA), which are both more volatile.
While the precise definition of the weight function does not matter much, we use this one, because it has 
only one single free parameter $T$ (which reduces the risk of overfitting the data), and because it 
can efficiently be computed recursively. It is similar to a moving average cross-over
(also shown in fig. 2, left), where 
the trend strength is proportional to the average price over a short time period $T/6$ minus the average price 
over a long time period $T$. To reduce the effect of outliers, we cap/floor the trend strength 
at $\pm 2.5$. We also cap/floor returns at 20 standard deviations from their mean when computing trend strengths.\\

\begin{figure}[t]\centering
        \includegraphics[height=5.2cm]{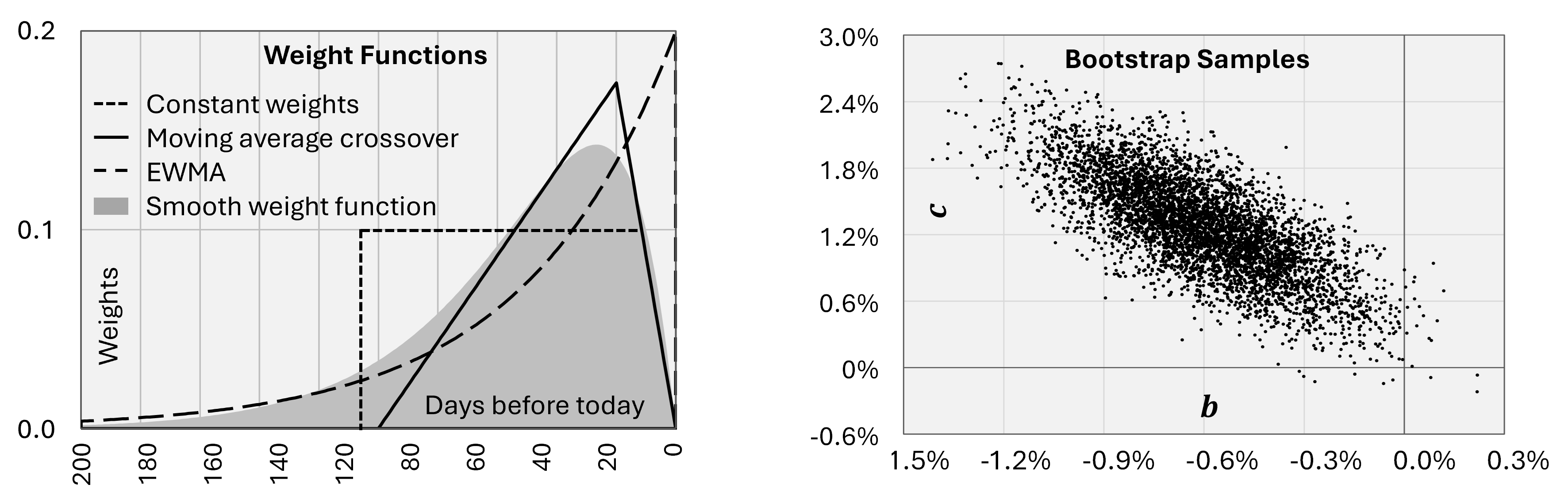}    
        \caption{{\it Left}: Different weight functions for weighting past returns in today's trend strength. We use the ``smooth weight function". {\it Right}: Bootstrap samples for the regression coefficients $b$ and $c$ in (\ref{cubic}). We read off their estimation errors from the widths of their distributions. 
}\label{figA}
\end{figure}

In addition to the trend strengths, 
we also need the variance $\sigma_{t}^2$ of each asset on day $t$. 
It is not constant but well-known to vary in time. Following the industry standard \cite{riskmetrics},
we measure the variance by an exponentially-weighted moving average (EWMA), defined as 
\begin{equation}
\sigma_{t}^2\ =\ (1-\alpha)\sum_{i=0}^\infty \alpha^i\cdot r_{t-i}^2\ =\ \alpha\sigma_{t-1}^2+(1-\alpha) r_{t}^2\ \ \ \text{with}\ \ \ \alpha=0.94.\label{EWMA}
\end{equation}
Here, we neglect the tiny mean of $r_{t}$ that we have modeled in (\ref{cubic}).\\

We then combine returns, volatilities, and trend strengths in a single data table,
a small extract of which is shown in table 2. This table is the basis of our analyses.
E.g., by regressing column 3 against the last column and its third power, 
the regression results (\ref{cubic}) can be obtained for a 1-year trend horizon.
The regression coefficients turn out to be remarkably
similar across different horizons. Therefore, it makes sense to also define ``aggregate factors"
as the averages of the $n$-th powers of the 8 trend strengths $\phi_k$ with horizons $T=2^k$ days:
\begin{eqnarray}
\phi_\text{agg}(t)&=&{1\over8}\big[\phi_1(t)+...+\phi_8(t)\big]\ \ \ ,\ \ \ 
\phi^{(n)}_\text{agg}(t)\ =\ {1\over8}\big[\phi_1^n(t)+...+\phi_8^n(t)\big].
\label{aggregate}\end{eqnarray}
By regressing column 3 against $\phi_\text{agg}$ and $\phi_\text{agg}^{(3)}$,
we obtain what we call the ``aggregate regression results", 
which apply the same factor loadings $b,c$ to all trend horizons. In our analyses, we will first 
compute the aggregate results, and then split them up by trend horizon.\vspace{2mm}

\begin{table}[h!]
\centering
{\small
\begin{tabular}{ |p{0.6cm}|p{1.5cm}|p{1.6cm}||p{1.6cm}||p{0.7cm}|p{0.7cm}|p{0.7cm}|p{0.7cm}|p{0.7cm}|p{0.7cm}|p{0.7cm}|p{0.7cm}|	}\hline
	\multicolumn{2}{|c|}{\it Database } &$ R_i(t+1)$&$\sigma^2_\text{EWMA}(t)$&\multicolumn{8}{|c|}{Trend strengths $\phi_T(t)$ for 8 time scales}\\	\hline
	Day& Market  &{\it Next}-day& $\alpha=0.94$&2d   &4d & 8d & 3w & 6w & 3m & 6m & 1y \\	\hline\hline
1	&S\&P 500&-0.2& 1.7&	0.3&0.7	&1.0&0.6	&	0.2	&	0.3	&	0.6	&	0.6	\\ \hline
1	&EUR/\$&-0.1& 0.6&0.2&0.2&0.0&-0.4&-0.6	&	-0.6	&	-0.8	&	-0.9\\ \hline
1	&Gold&-0.5& 1.1&-0.3	&-0.7&-0.7&0.1	&	1.1	&	1.5	&	1.1	&	0.4	\\ \hline\hline
2	&S\&P 500&-0.3&1.5 &-0.1	&0.4&	0.9	&	0.7	&	0.2	&	0.3	&	0.6	&	0.6\\ \hline
2	&EUR/\$&-1.0& 0.8&-0.7&-0.2&-0.2	&-0.4&-0.6	&-0.6	&-0.8&	-0.9\\ \hline
2	&Gold&0.8& 1.1&0.3&-0.2	&-0.6	&	0.1	&	1.1	&	1.5	&	1.1	&	0.4\\ \hline
\end{tabular}}
\caption{Small extract of the data table used in our analysis. Only two of the 9216
business days and only three of the 24 markets are shown. 
Column 3 shows the normalized daily log-returns, 
whose long-term variance is 1. Column 4 shows their 
shorter-term EWMA variance. The other columns contain the 8 trend strengths on the given day for the given market.}
\label{tab:all_var_reg}
\end{table}

\subsection{Statistical Analysis}

Most of our analyses are based on ordinary least squares (OLS) regression.
In the case where we want to predict the variance $\hat\sigma_i^2$ of 
returns $r_i$ with $i\in\{1, ..., N\}$, this
minimizes the mean squared error MSE of predicted values $\hat\sigma^2$ and realized values $r_i^2$. 
We then report the resulting $R^2$:
\begin{equation}
\text{MSE}={1\over N}\sum_i (r_i^2-\hat\sigma_i^2)^2\ \ \ ,\ \ \ 
R^2=1-{\text{MSE}\over \text{var}(r^2)}. 
\label{ols}\end{equation}
We also perform Gaussian quasi-maximum likelihood estimation (QMLE) in some cases.
The negative quasi-log-likelihood for a univariate forecast $\hat\sigma_i^2$ is
\begin{equation}
\text{QLIKE}=\sum_i \Big[\log(\hat\sigma_i^2)+{r_i^2\over\hat\sigma_i^2}\Big].
\label{mle}\end{equation}
In this case, we report QLIKE per observation, QLIKE$/N$.
We prefer OLS minimization over QMLE in this paper, because OLS places greater weight on
observations with high volatility, which are the most important ones for risk management.\\

Given the small values of many of our regression coefficients, 
and the huge amount of noise in financial market data, 
we must be very careful in assessing the statistical significance of the results. 
Since market returns cannot be assumed to be identically distributed normal random variables, we cannot trust standard formulas for the t-statistics, adjusted R-squared's, and other statistics.\\

Instead, the test statistics reported in this paper
are measured empirically as follows.
The error of all regression coefficients ($a,b,c,...$) and their t-statistics are computed by bootstrapping: we draw 100 or more samples with replacement from the 9216 trading days covered by our analysis,
two of which are shown in table 2. 
Regression on these $\ge$100 bootstrap samples of days yields a distribution of $\ge$100 regression coefficients.
This is illustrated in fig. 2 (right) at the example of the coefficients 
$b$ and $c$ in (\ref{cubic}).
The estimation errors that we report are half of the difference between the 84th and the 16th 
quantile of this distribution.
These actual errors are typically 3-4 times as large as the errors reported by
standard regression tools, which significantly under-estimate them. \\

Drawing bootstrap samples of our days must not destroy the serial dependence of the time series and thereby
the trends, variance clusters, and cross-correlations of market returns. 
For the regression analyses of section 3, this is not an issue, because 
we first compute the trends and variances on each day, combine them with the {\it next}-day returns
in the same lines of table 2, and only afterwards resample the complete 1-day blocks in table 2. 
To preserve cross-correlations, each bootstrap sample includes the same days for each asset.
However, this procedure cannot be applied to the GARCH and stochastic volatility models of section 4, 
where variances and correlations must be computed day by day for each sample. 
For these models, we therefore use block-bootstrapping:
we split the data into 18 contingent blocks of 512 days each, and resample the whole blocks. \\

The adjusted R-squared is computed by 16-fold cross validation: 
we split our 9216-day time period into 16 adjacent 576-day periods.
We then combine 15 of these sub-periods into a training sample.
This yields an out-of-sample prediction of the target variable for the 16th period (the validation sample). 
Repeating this 16 times such that each sub-period appears once as the validation sample,
we get an out-of-sample prediction for the target variable over the whole period. 
Our out-of-sample R-squared $R^2_{adj}$ is obtained from the MSE of this predicted target variable and the actual target variable (the results do not change significantly if we use only training periods {\it before} the validation period).
The estimation error of the R-squared's is computed by 
block-bootstrapping of the 16 sub-periods. \\

Analogous procedures are used for the out-of-sample QLIKE in the case of QMLE.

\section{EWMA-based Volatility Model}
\label{sect:var}

\begin{figure}[t!]\centering
    \includegraphics[height=5.5cm]{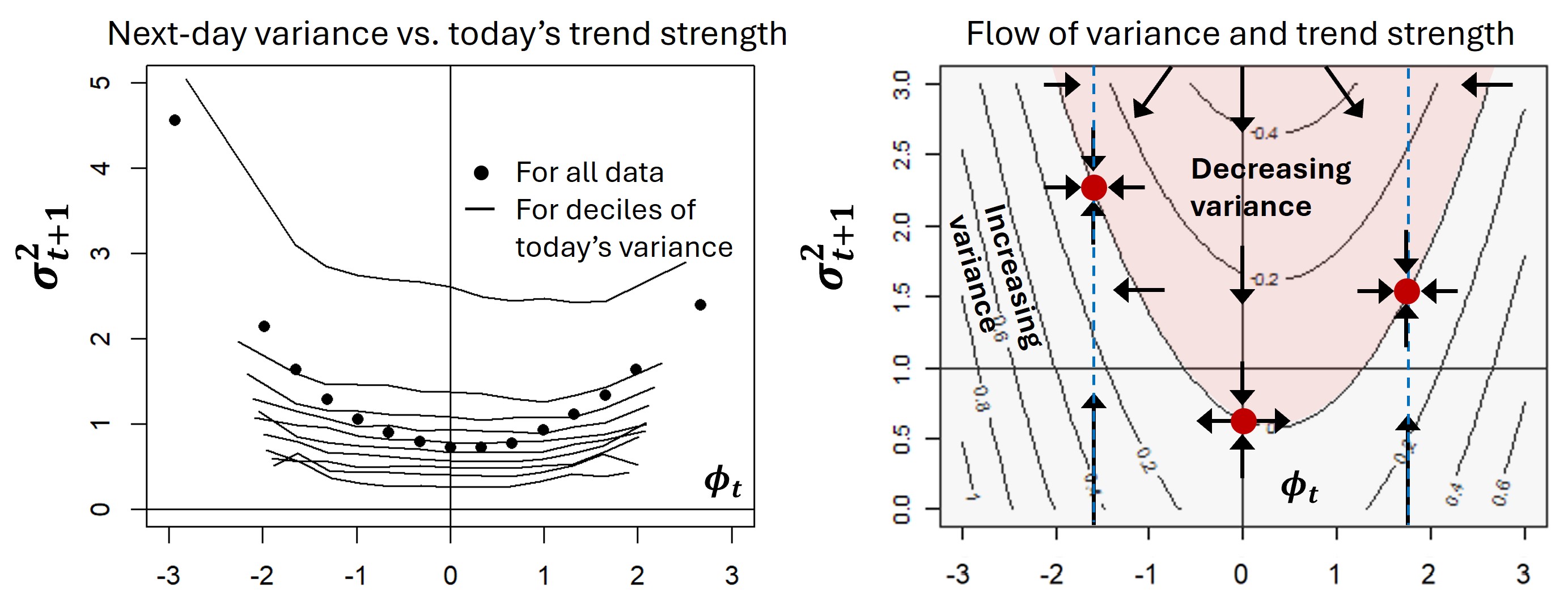}
    \caption{Left: Tomorrow's variance $\sigma_{t+1}^2$ as a function of today's trend strength $\phi_t$, both overall (dots) and for deciles of the variance (lines).
    Right: contour plot of the mean daily change of the variance as a function of the 
    current trend strength (horizontal axis) and the current variance (vertical axis).
   Arrows indicate the changes of variance and trend strength. }
    \label{fig:var_k}
\end{figure}

We will model the future volatility of returns or its square, the variance, as a function of the current trend strength,
aggregated across our 24 markets, 8 time scales, and 35 years of data.
We proceed in analogy with model (\ref{cubic}) of future returns.\\

A new element is that we must also include the current variance as a regression factor, 
as it is an important explanatory variable for the future variance. 
In fact, it turns out to be the most important factor by far.
Our task is then to quantify by how much its predictive power can be enhanced by adding trends as 
additional factors.
In this section, we do this for a basic model, in which we measure the 
current variance as the EWMA variance (\ref{EWMA}).  
In section 4, we will generalize the analysis to more advanced volatility models.

\subsection{Visualization of the Data}

We first visualize our data by grouping them into 15 bins $b_k$ of similar trend strength:
\begin{equation}
\phi\in b_k\ \ \ \text{with}\ \ \ k\in\{-7,+7\},\ \ \ b_k=[\ {k\over3}-{1\over6}\ ,\ {k\over3}+{1\over6}\ [\ .\notag
\end{equation}
I.e., bucket $b_0$ covers trend strengths from $-1/6$ to $+1/6$, $b_1$ from $+1/6$ to $+1/2$, and so on.
We extend the outermost buckets $b_{\pm7}$ to $\pm\infty$.
We then measure the variance of next-day returns $r_{t+1}$, given that the trend strength lies in bin $b_k$ on day $t$.
For daily data, the dots in fig. 1 (right) and in fig. 3 (left) show this variance for each
bin as a function of the average trend strength in that bin.
The error bars in fig. 1 (right) are the estimation errors for each bin.
The results are aggregated across the 8 different trend horizons $T\in\{2,4,8,...,256\}$ days. 
The figures suggest to model tomorrow's variance by an at least second-order polynomial of today's trend strength. 
This polynomial appears to have a negative linear component: the parabola in fig. 3 (left)
looks like a skewed "smile".\\

\begin{figure}[t!]\centering
    \includegraphics[height=5cm]{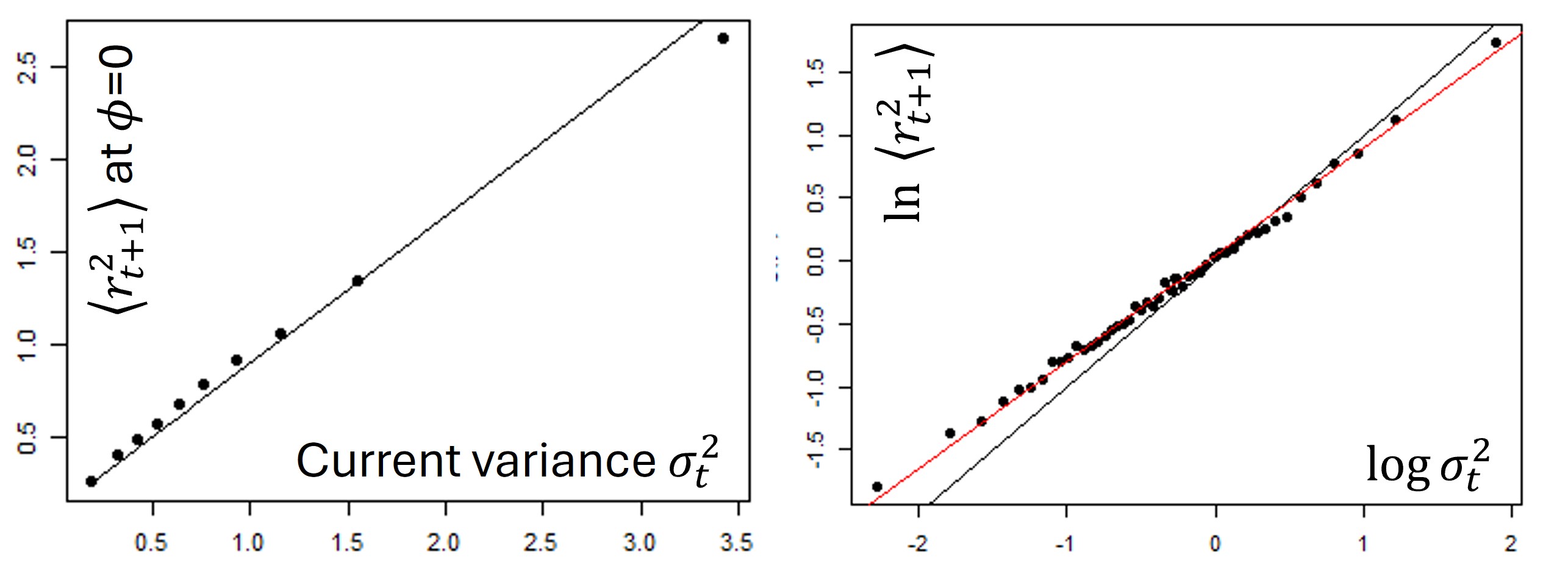}
    \caption{Left: The next day square-return (vertical axis) appears to be an almost linear
    function of today's variance $\sigma^2$ (horizontal axis).
    Right: However, a log-log plot suggests that it could also be
    a power slightly below 1 (slope of the red line) of today's variance.}
    \label{fig:var_k}
\end{figure}

The variance $\sigma_t^2$ of financial market returns on day $t$
is well-known to be auto-correlated and to come in clusters.
As discussed, we therefore also include today's EWMA variance $\sigma_t^2$ (\ref{EWMA}) as an 
explanatory variable for tomorrow's variance. 
Fig. 3 (left) repeats the above analysis for each decile of the past 60-day variance (solid lines). 
The dependence of tomorrow's variance on today's 
trend strength still looks like a skewed smile, but with much smaller curvature.
In addition, there is a dependence on today's variance, which grows monotonously with today's trend strength.
Apparently, much of the quadratic dependence of tomorrow's variance on today's trend strength is already contained in today's variance.\\

Fig. 4 (left) plots the average of tomorrow's square return $\langle r_{t+1}^2\rangle$ (as an estimate of the future variance) against today's variance $\sigma_t^2$. 
At first sight, $\langle r_{t+1}^2\rangle$ seems to be an almost linear function of $\sigma_t^2$. 
However, a more fine-grained log-log plot (fig. 4, right) 
suggests that $\langle r_{t+1}^2\rangle$ could alternatively 
be modeled as a power $(\sigma_t^2)^\kappa$.  
We will discuss the linear model in this section, and return to the 
alternative power-law ansatz in subsection 4.3. in the context of stochastic volatility models,
where it naturally arises.\\

Fig. 3 (right) shows a contour plot of the change of the variance $\langle r_{t+1}^2-\sigma_t^2\rangle$
as a function of both the trend strength (horizontal axis) and today's variance (vertical axis).
Arrows indicate the changes of the variance and the trend strength;
it follows from (\ref{cubic}) that the trend strength is driven to $\phi=\sqrt{-b/c}\approx\pm1.5$, where
the expected next-day excess return over the long-term risk premium is zero.
This yields a fixed line in the flow diagram of fig. 3 (right) with 2 stable and 1 unstable fixed points.
In times of strong trends, the variance tends to increase day after day towards the fixed line.
This explains why the variance is high after strong trends have been developing: it is natural to 
identify the fixed line in fig. 3 (right) with the parabola of dots in fig 3 (left).

\subsection{Regression Analysis}

We now quantify these qualitative observations. We start with a
simple linear EWMA-based variance model analogous to (\cite{riskmetrics, mincer}) that 
predicts tomorrow's variance $\sigma_{t+1}^2$ from today's variance $\sigma_t^2$ and the long-term variance $\sigma_0^2$:
$$\sigma^2_{t+1}=\lambda\sigma_0^2+(1-\lambda)\sigma_t^2,$$
such that the variance reverts to its equilibrium value $\sigma_0^2$ within a time span of order $1/\lambda$ trading days. In our case, $\sigma_0^2=1$. 
Fig. 3 (left) suggests to refine this model by taking not only the 
current variance, but also the current trend strength $\phi_t$ into account.
We thus include linear and quadratic powers of today's trend strength on the right-hand side, and 
estimate $\sigma_{t+1}^2$ on the left-hand side by the next-day squared return $r_{t+1}^2$:
\begin{equation}
 r_{t+1}^2=a+d\sigma_t^2+e\phi_{\text{agg},t}+f\phi_{\text{agg},t}^{(2)}+\epsilon_{t+1}, \label{volatrend}
\end{equation}
where $\epsilon$ represents the noise. 
Here, we use the aggregate trend strength (\ref{aggregate})
to include all trend horizons.
We have also tried out other functions of the trend strength, such as trigonometric functions, $n$-th roots, etc.,  but the second order polynomial regression 
has yielded the best out-of-sample (or ``adjusted") R-squared.

\begin{table}[h!]
\centering
\begin{tabular}{ |p{4cm}||p{1.5cm}|p{1.5cm}||p{1.5cm}|p{1.5cm}||p{1.5cm}|p{1.5cm}|}	\hline
	{Variance Model}& \multicolumn{2}{|l||}{Trends \& Variance} & \multicolumn{2}{|l||}{Only Trends}& \multicolumn{2}{|l|}{EWMA Variance} \\	\hline
	{Coefficient}& Value  &t-stat. & Value  &t-stat.  & Value  &t-stat. \\	\hline
	$a$ &$ +0.05 $ & 1.4 &$+0.24$& 11  &$+0.16$ & 3.6 \\ 
	$d$ &$ +0.67$ & 10 &-  &-  &$+0.84$&16 \\
    $e$ &$ -0.21$ & 8.5 &$-0.36$&11 &- &- \\	
	$f$ &$ +0.34$ & 10 &$+0.91$& 23&- &- \\	\hline
    MSE&8.662& 4.5 & $9.096$&  3.7  & 8.721& 4.5\\ 
    MAE&1.059& 14 & $1.094$&  16  & 1.063& 14\\ \hline
    $R^2\cdot10^4$ &1144& 7.7 & $738$&  9.3  & 1072& 9.4\\ 
	$R^2_{adj}\cdot 10^4$&1127& 6.5   & $682$&  5.1  & 1067& 28\\ 
   	$\Delta R^2_{adj}\cdot 10^4$&60& 7.5   & -&  - & -& -\\ \hline
\end{tabular}
\caption{Regression results for the next-day variance for model (\ref{volatrend}) (left),
for a pure trend model without $d$ (center), and for a pure variance model
without $e$ and $f$ (right).
The estimation errors and t-statistics are measured empirically from 300 bootstrap samples. 
The out-of-sample R-squared $R^2_{adj}$ is computed by 16-fold cross validation. The last line reports
the increase of $R^2_{adj}$ of the "Trends and Variance" model over the "EWMA Variance" model.}
\label{tab:all_var_reg}
\end{table}

Table 3 shows the regression results. 
We observe that the variance is the most significant factor, with the difference $d-1\approx-0.2$ measuring how strongly
the variance tends to revert to its equilibrium value. However,
the linear ($e$) and quadratic ($f$) dependencies on the trend strength are also highly significant.
$f$ is positive, which explains why the variance keeps growing during strong trends.
$e$ is negative, which reflects the asymmetry of fig. 3 (left). 
Given that trends measure cumulative recent returns, this is in line with
the so-called ``leverage effect" that is well-known in equity markets: 
strong negative returns tend to trigger a subsequent rise of volatility.\\

For comparison, table 3 also shows the regression results with only trends or only variance as explanatory factors. Clearly, all three factors are needed to optimize the out-of-sample $R^2$.
The same is true for the out-of-sample mean-squared error (MSE) or the mean absolute error (MAE).
The estimation errors and thereby the $t$-statistics of the out-of-sample $R^2_{adj}$
and the MSE/MAE are estimated from 300 block-bootstrap samples. 
The last line shows how much $R^2_{adj}$ one gains by adding trends as regression factors,
compared with using only the variance.
The gain of $0.006$ may seem small, but it is
statistically highly significant. The economic significance is significant
for predictions longer than 1 day (see subsection 3.6).

\subsection{Maximum Likelihood Estimation}

Table 3 is based on ordinary least squares (OLS) regression.
For comparison, the results of a quasi maximum likelihood estimation (QMLE) of the
parameters are shown in table 4.

\begin{table}[h!]
\centering
\begin{tabular}{ |p{4cm}||p{1.5cm}|p{1.5cm}||p{1.5cm}|p{1.5cm}||p{1.5cm}|p{1.5cm}|}	\hline
	{Maximum Likelihood}& \multicolumn{2}{|l||}{Trends and variance} & \multicolumn{2}{|l||}{Only trends}& \multicolumn{2}{|l|}{EWMA variance} \\	\hline
	{Estimation}& Value  &t-stat. & Value  &t-stat.  & Value  &t-stat. \\	\hline
	$a$ &$ +0.110 $ & 15 &$+0.384$& 58  &$+0.104$ & 15 \\ 
	$d$ &$ +0.717$ & 44 &-  &-  &$+0.914$&67 \\
    $e$ &$ -0.072$ & 9.0 &$-0.201$&16 &- &- \\	
	$f$ &$ +0.209$ & 18 &$+0.703$& 49&- &- \\	\hline\hline
    1-QLIKE/N &0.2532& 29  & 0.1872&  17   & 0.2448& 26\\ 
	1-QLIKE/N (o.o.s.) &0.2528& 3.9 & 0.1853&  3.1  & 0.2445& 4.1\\ 
   	$\Delta$ QLIKE/N (o.o.s.)&0.0084& 8.7 & -& - & -& -\\ \hline
\end{tabular}
\caption{Maximum likelihood estimation for the same models as in table 3. 
}
\label{tab:all_var_reg}
\end{table}

Here we report 1-QLIKE/$N$,
which is zero for a constant variance benchmark model, and which we want to maximize.
We again observe that this measure is increased by adding trends as explanatory factors.
The increase for next-day predictions is small, but statistically highly significant. Note that 
the trend coefficients $e$ and $f$ are smaller than for the OLS regression analysis. 
This has a transparent explanation. OLS regression minimizes the mean square error  (\ref{ols})
of the predicted variance $\hat\sigma_{t+1}^2$ on day $t+1$ relative to the realized variance $r_{t+1}^2$.
With a factor ansatz such as (\ref{volatrend}), differentiating with respect to the factor loadings $\vec\beta$ yields the condition
$$\hat\sigma^2_{t+1}=\alpha+\vec\beta\cdot \vec f_t,\ \ \ \Rightarrow\ \ \ 
\sum_t \vec f_t(r^2_{t+1}-\alpha-\vec\beta\cdot \vec f_t)=0,$$
where $\vec f$ is the vector of factors. 
By contrast, Gaussian QMLE minimizes the quasi-maximum likelihood loss (\ref{mle}).
Differentiating with respect to $\vec\beta$ now yields the condition
$$\sum_t w_t^2\cdot \vec f_t(r^2_{t+1}-\alpha-\vec\beta\cdot \vec f_t)=0,\ \ \ \text{with}\ \ \ w_t= {1\over\hat\sigma^2_{t+1}}$$
$w_t$ can be interpreted as the weights of a weighted OLS regression
(since the weights themselves depend on the predicted variance, this equation must be solved iteratively).
Thus, days with high predicted variance 
$\hat\sigma^2_{t+1}$ have less weight in QMLE. But those are the days
on which trends tend to be strong and, in particular, the leverage effect is important.
This weighting can thus reduce the significance of the trend coefficients.
Since accurate volatility forecasts are most-needed in crises, when volatility is high, 
we do not consider this a disadvantage of OLS regression, on which we will primarily focus in the following.

\subsection{Refinements by Trend Horizon, Asset Class and Time Period}

\begin{figure}[t!]\centering
    \includegraphics[height=5cm]{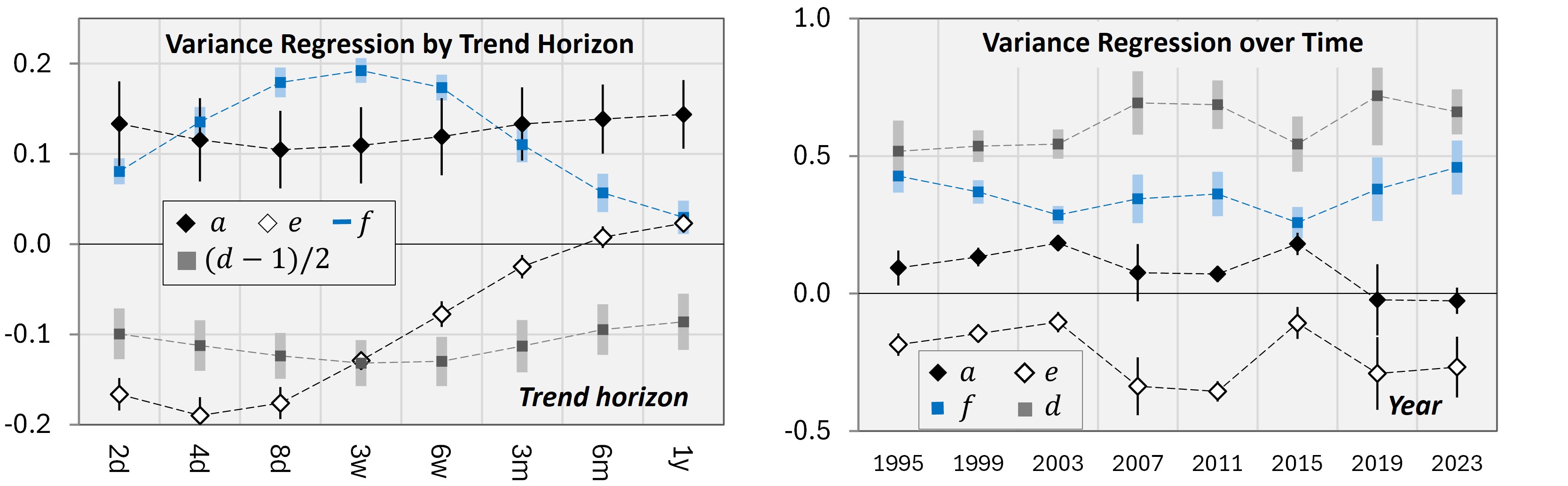}
	\caption{{\it Left:} variance regression coefficients as a function of the trend horizon.
    Only trends with horizon up to 3 months can help to predict the variance. {\it Right:} coefficients for an 8-year rolling time window. 
    They are relatively stable in time within error bars.}
    \label{fig:var_k}
\end{figure}

The regression results of table 3 are aggregated over trend time horizons $T=2^k$ days 
with $k\in\{1, ..., 8\}.$
In a second step, we repeat this analysis for each $k$ individually (fig. 5, left).
One observes that the coefficients related to the trend strength ($e,f$) become 
insignificant for $T\ge 2^7$ business days $\approx 6$ months. Thus, only trends with horizons up to 3 months have 
significant predictive power for the future variance. 
The linear term, which captures the leverage effect, mostly stems from even shorter trend horizons
from 1 day up to about a month. It becomes insignificant for horizons of 3 months or more.\\

To check in how far the results depend on the asset class, we have repeated the analysis separately 
for equity indices, interest rates, currencies and commodities.\\

The results are shown in table 5. One observes that the results for interest rates, 
currencies, and commodities are universal within estimation errors,
so we may apply the same volatility model to all of them.
However, equities differ in that they
display a much stronger linear dependence $e$ on the trend strength. 
In fact, each of our six equity indices turns out to yield the same 
high value of $e$ within error bars. This confirms that
the leverage effect is mostly an equity phenomenon.
Equities also tend to have a higher quadratic trend coefficient $f$ and - as a result 
of the higher $e$ and $f$ -
a higher out-of-sample $R$-squared.\vspace{3pt}

\begin{table}[h!]
\centering
\begin{tabular}{ |p{2.2cm}||p{1.5cm}|p{1cm}||p{1.5cm}|p{1cm}||p{1.5cm}|p{1cm}||p{1.5cm}|p{1cm}|}	\hline
	{Asset Class}& \multicolumn{2}{|l||}{Equities} & \multicolumn{2}{|l||}{Interest rates}& \multicolumn{2}{|l||}{Currencies}& \multicolumn{2}{|l|}{Commodities}\\	\hline
	{Coefficient}& Value  &Error & Value  &Error  & Value  &Error & Value  &Error  \\	\hline
$a$ & $+0.06$ &  0.05 & $+0.13$ &0.03  & $+0.10$ & 0.05  & $+0.03$ & 0.07 \\ 
$d$ & $+0.62$ & 0.09   & $+0.62$  &0.05   & $+0.65$ & 0.08  & $+0.66$ & 0.13 \\
$e$   &$-0.60$ &  0.05   & $-0.10$ &0.03  & $-0.10$ & 0.03  & $-0.11$ & 0.05 \\
$f$  & $+0.46$ &  0.08  & $+0.27$  &0.03  & $+0.28$ & 0.04  & $+0.37$ & 0.07 \\ \hline
$R^2_{adj}$ & $+0.17$ &  0.03  & $+0.10$  &0.02  & $+0.09$ & 0.02  & $+0.08$ & 0.01 \\ \hline
\end{tabular}
\caption{Variance regression results for individual asset classes. The coefficient for interest rates, currencies and commodities agree within estimation errors, while equity indices
have a much stronger linear trend dependence $e$, reflecting the leverage effect.}
\label{tab:trends_var}
\end{table}

We have also tested to what extent our regression coefficients 
have changed over time. To this end, we have repeated the analysis of table 3 for the ``trends and variance" model  
for an 8-year rolling window, where two adjacent rolling windows overlap by 4 years.
Fig. 5 (right) shows the resulting regression coefficients. 
As one can see, they are relatively stable in time
in the sense that they agree within their estimation errors.
There are no obvious long-term trends of these coefficients.

\subsection{Enhancement of the Model}

A closer look at fig. 3 (left, top line) reveals that its asymmetry, which reflects the leverage effect,
is more pronounced when the variance is high. To test this, we generalize ansatz (\ref{volatrend}) to also include products of the trend strength $\phi_t$ and the current variance $\sigma_t^2$ as factors:
\begin{equation}
r_{t+1}^2=a+d\sigma_t^2+e_1\phi_{\text{agg},t}+e_2\sigma_t^2\phi_{\text{agg},t}+f_1\phi_{\text{agg},t}^{(2)}
+f_2\sigma_t^2\phi_{\text{agg},t}^{(2)}+\epsilon_{t+1}.
\label{extend1}\end{equation}
The regression results are shown in table 6 (left columns).
The out-of-sample $R^2_{adj}$ is significantly higher than that of 
the original model in table 3.
$f_2$ is insignificant, so we can drop it.
On the other hand, $e_2$ is highly significant.
Within estimation errors (values divided $t$-statistics), $e_2\approx-e_1$.
We therefore simplify ansatz (\ref{extend1}) to
\begin{equation}
 r_{t+1}^2=a+d\sigma_t^2+e\phi_{\text{agg},t}\cdot(\sigma_t^2-1)+f\phi_{\text{agg},t}^{(2)}+\epsilon_{t+1}. 
\label{extend2}\end{equation}
As seen from table 6 (center columns), this simplification slightly decreases the in-sample $R^2$,
but it further increases the out-of-sample $R^2$.
The simple enhancement (\ref{extend2}) of our original model (\ref{volatrend})
captures the asymmetry observed in fig. 3 (left). The
linear trend coefficient $e$, which reflects the leverage effect, is now multiplied by a 
positive number in times of above-average variance, and by a negative number in times of below-average variance.
\\

We can further enhance the model based on what we have observed in fig. 5:
trends with horizons of 6 months or longer are irrelevant for predicting the variance.
Therefore we only include time horizons $T=2^k$ days with $k\in\{1, ... ,6\}$
in the aggregate trend strength (\ref{aggregate}). This further improves the adjusted $R$-squared,
as shown in table 6 (right columns).

\begin{table}[h!]
\centering
\begin{tabular}{ |p{2.7cm}||p{1.5cm}|p{1cm}||p{2.3cm}||p{1.5cm}|p{1cm}||p{1.5cm}|p{1.1cm}|}	\hline
	{Enhancement}& \multicolumn{2}{|l||}{Step 1} &&\multicolumn{2}{|l||}{Step 2}
    &\multicolumn{2}{|l|}{Step 3}\\	\hline
	{Coeff.}& Value  &t-stat &Coeff. & Value &t-stat & Value &t-stat\\	\hline
	$a$      &$ +0.20$   & 4.9 &$a$&$ +0.19$ &10 &$+0.20$&9.7 \\ 
	$d$      &$ +0.39$   &6.4  &$d$&$ +0.40$ &12&$+0.44$&13 \\
    $e_1$ &$ +0.23$&3.8&$e$&$-0.31$&5.1 &$-0.29$&5.9 \\	
	$e_2$    &$ -0.30$   &6.2  &-&-        &-    &-    &-  \\	
    $f_1$ &$ +0.43$&12&$f$&$+0.43$ &14   &$+0.39$&13 \\	
	$f_2$    &$ +0.00$   &0.1  &-&-        &-    &-    &-   \\	\hline
    $R^2\cdot10^4$ &1263     &10 & $R^2\cdot10^4$&1261    &10   &1305 &11\\ 
	$R^2_{adj}\cdot10^4$ &1219&4.9 &$R^2_{adj}\cdot10^4$&1225    &4.8   &1264 &4.8\\ 
   	$\Delta R^2_{adj}\cdot10^4$ &92&2.1&$\Delta R^2_{adj}\cdot10^4$&98&2.1   &137  &2.2 \\ \hline
\end{tabular}
\caption{Enhancement of the model by variance-dependent trend terms.
In step 1, all terms linear in $\sigma^2$ are included. 
In step 2, the insignificant term $f_2$ is excluded and 
$e_1$ and $e_2$ are combined. In step 3, the irrelevant trend horizons $T>3$ months
are removed. The last line shows the improvement of the out-of-sample $R^2_{adj}$ 
relative to the original model (\ref{volatrend}).}
\label{tab:all_var_reg}
\end{table}

\subsection{Longer-term Predictions}

In view of the very strong parabolic dependence of tomorrow's variance on today's trend 
strength observed in fig. 1 (right), it may at first be surprising that the additional $R$-squared 
achieved by trends is relatively small. 
The reason is that our regression analyses also contain today's variance.
As we have seen, the variance tends to increase 
day after day in times of significant trends. This increase accumulates over weeks or months, depending on the duration of the trend. 
The parabola of fig. 1 (right) reflects this cumulative increase. 
However, most of this cumulative increase is already contained in today's variance,
which is therefore the dominant factor when predicting the next-day variance. \\

The situation changes when we try to predict
the next-week, next-month, or longer-term variance from today's variance and trends. 
This is relevant, e.g., for investment portfolios that are rebalanced on a weekly, monthly, 
or quarterly basis, based on expected returns and variances.
For a rebalancing horizon of $T=2^{k-1}$ days, one should use as explanatory factors 
only trends with trend horizon $T\ge2^k$. Likewise, the EWMA variance should take
the rebalancing horizon into account. We will use $\alpha=0.94$ in (\ref{EWMA}) for $k\le3$, 
and $\alpha=1-2^{-(k+1)}$ for $k>3$, corresponding to half-lifes $-\ln 2/\ln\alpha$ 
of $11$ resp. $1.4\cdot2^k$ days.
For $k=1$, this recovers our previous analyses.
These values of $\alpha$ also turn out to ensure that the changes of the EWMA variance and 
of the shortest trend strength, which both contribute to the rebalancing costs,
are of the same order of magnitude.\\

\begin{table}[h!]
\centering
\begin{tabular}{ |p{1.6cm}||p{0.9cm}|p{0.7cm}|p{0.9cm}|p{0.7cm}|p{1.1cm}|p{0.7cm}|p{0.9cm}|p{0.7cm}||p{1cm}|p{1cm}|p{1cm}|}	\hline
{Period}& a   &\ $t_a$&d &\ $t_d$& e    &\ $t_e$&f &\ $t_f$&$R^{2,\text{EW}}_{adj}$&$R^{2}_{adj}$&$\Delta R^2$\\ \hline 
1-day   &0.20  &9.7 &0.44 &13 &$-0.29$ &5.9 &0.39 &13  &0.107 &0.126 &+18\%\\
2-day   &0.21  &9.8 &0.43 &13 &$-0.23$ &6.4 &0.40 &14   &0.168 &0.196 &+17\%\\
1-week  &0.23  &8.7 &0.46 &12 &$-0.15$ &5.2 &0.34 &15   &0.231 &0.265 &+14\%\\
2-week  &0.23  &6.5 &0.48 &11 &$-0.13$ &3.8 &0.32 &13   &0.233 &0.280 &+20\%\\
1-month &0.26  &5.0 &0.49  &9.3 &$-0.12$ &3.2 &0.28 &11   &0.203 &0.268 &+32\%\\
2-month &0.30  &4.3 &0.48  &7.4 &$-0.10$ &1.8 &0.24 &10  &0.168 &0.238 &+41\%\\ 
3-month &0.34  &3.8 &0.46  &5.5 &$-0.09$&1.1 &0.22 &10 &0.115 &0.176 &+53\%\\ \hline
\end{tabular}
\caption{Regression results for the next-$T$-day variance instead of the next-day variance
for various prediction horizons $T$. The last three columns show
the out-of-sample $R$-squared's for (i) the EWMA-based model without trends and (ii) the enhanced model (\ref{extend2}), as well as the improvement of (ii) over (i) in \%. The improvement grows with the prediction horizon.}
\label{tab:trends_var}
\end{table}

Table 7 shows the regression results for predicting the $T$-day variance  
for the refined model of subsection 3.5. They were computed for $T=2^k$ days 
and then linearly interpolated to 5, 10, 22, and 44 days.
All coefficients except $e$ remain highly significant up to a 3-month prediction horizon. 
In fact, they do not depend much on the horizon. 
The linear trend coefficient $e$, which represents the shorter-term leverage effect,
naturally decreases for longer prediction horizons. 
The increase in $R^2$ of 
the "Trend+Variance" model over the linear EWMA-based model
is indeed larger for longer trend horizons.
Thus, while the economic value of adding trends as explanatory factors may be small
for next-day variance predictions, it can certainly be economically significant for 
longer-term predictions. 

\section{Advanced Volatility Models}

In section 3, we have illustrated at a simple linear EWMA-based regression model
how variance forecasts can be improved by including medium-term trends
as explanatory factors.
In this section, we investigate more briefly whether accounting for trends can 
similarly improve more advanced volatility/variance models. \\

Out of the many variance models, 
we select as key representatives the GARCH(1,1) and GJR-GARCH models, 
a GARCH(1,1) model based on price ranges, the HAR model, the Heston model, and
the Leverage stochastic volatility (SV) model.
We find that adding medium-term trends as explanatory factors can moderately, 
but statistically highly significantly improve the 1-day forecasts of 
all of these models. 

\subsection{GARCH(1,1) and GJR-GARCH Models}

We begin with the GARCH(1,1) model \cite{garch} as a benchmark:
$$r_{t}=\sigma_{t}\epsilon_{t},\ \ \ \sigma_{t+1}^2=\omega+\alpha r_t^2+\beta\sigma_t^2,\ \ \ E(r_{t}^2)\sim\sigma_{t}^2$$
where the GARCH variance $\sigma_t^2$ is computed iteratively day by day.
We extend the second equation to also include trends based on the refined model of subsection 3.5:
\begin{equation}
\sigma_{t+1}^2=\omega+\alpha r_t^2+\beta\sigma_t^2+\lambda(\sigma_t^2-1)\phi_{\text{agg},t}+\zeta\phi_{\text{agg},t}^{(2)}.
\label{garch11}\end{equation}
Table 8 (left columns) shows the regression results with and without trends for our
35 years of daily returns across 24 diversified markets.
The trend coefficients $\lambda$ and $\zeta$ are statistically highly significant.
The improvement of the GARCH(1,1) model with trends ($R^2_{adj}=0.1219$)
over the GARCH(1,1) model without trends ($R^2_{adj}=0.1121$) is also highly
significant. Note that the increase by $0.0098$ (for 1-day variance forecasts)
is bigger than the increase by $0.0054$ of the established GARCH(1,1) model 
over the simple linear EWMA-based variance model of section 3.5, for which 
$R^2_{adj}=0.1067$ without trends (based on the same data set). 
We conclude that including trends as explanatory factors can also improve 
GARCH(1,1) volatility forecasts, just like it can improve the EWMA volatility forecasts.

\begin{table}[h!]
\centering
\begin{tabular}{ |p{2.5cm}||p{1.5cm}|p{1cm}|p{1.5cm}|p{1cm}||p{1.5cm}|p{1cm}|p{1.5cm}|p{1cm}|}	\hline
	{}& \multicolumn{2}{|l|}{GARCH} & \multicolumn{2}{|l||}{With trends}& \multicolumn{2}{|l|}{GJR-GARCH}& \multicolumn{2}{|l|}{With trends}\\	\hline
	All assets  & Value  &t-st.  & Value  &t-st. &Value  &t-st.&Value  &t-st.\\	\hline
$\omega$       &$+0.029$&2.4&$+0.096$&3.9&$+0.031$&3.4&$+0.071$&2.9\\
$\alpha$        &$+0.107$&4.5&$+0.114$&5.5&$+0.035$&4.2&$+0.024$&2.1\\
$\beta$         &$+0.865$&24 &$+0.679$&10 &$+0.875$&31 &$+0.746$&13\\
$\lambda$       &-      &-  &$-0.095$&3.1&-    &-  &$-0.066$&2.5\\
$\zeta$          &-      &-  &$+0.133$&3.6&-    &-  &$+0.089$&4.1\\
$\gamma$       &-      &-  &-       &-  &$+0.115$&3.4&$+0.133$&3.7\\ \hline\hline
$R^2\cdot10^4$ &$1163$ &6.3&$1274$  &6.8&1265 &5.5&$1355$  &5.7\\
$R^2_{adj}\cdot10^4$&$1121$&5.8&$1219$  &6.3&1210 &5.5&$1294  $&5.7\\  
$\Delta R^2_{adj}\cdot10^4$&  -  &  - &$98$  &3.6& -    & -  &$83 $  &5.1\\  \hline
\end{tabular}
\caption{Adding medium-term trends as explanatory variables to a GARCH(1,1) model 
or to a GJR-GARCH model can improve their predictive power.  
The last line shows the improvement of the out-of-sample $R^2$ (as measured by cross-validation)
relative to the models without trends, as well as its $t$-statistics (as measured by block-bootstrapping).}
\label{tab:all_var_reg}
\end{table}


The GARCH(1,1) model does not capture the asymmetric leverage effect 
discussed in section 3
(large negative returns increase the future volatility more than large positive returns).
A popular refinement that does is the GJR-GARCH model \cite{gjr}:
$$r_{t}=\sigma_{t}\epsilon_{t}\ \ \ ,\ \ \ \sigma_{t+1}^2=\omega
+(\alpha +\gamma I_{r_t<0})r_t^2+\beta\sigma_t^2,$$
where $I_{r_t<0}$ is 1 when $r_t$ is negative, and zero otherwise. We extend the GJR model
to trends by adding
the same factors as in (\ref{garch11}).
The results are also shown in table 8 (right columns). Our conclusion is the same as for
the GARCH model and the EWMA-based model: Accounting for trends can significantly improve the
predictive power of the GJR GARCH model.\\

Our reported out-of-sample $R$-squared's of 10-13\% seem relatively small. 
However, here we have aggregated across all assets and 35 years
to compare models with and without trends for as many data points as possible.
For sub-periods or sub-sets of assets, $R^2_{adj}$ can be higher. 
E.g., for equities over the whole 35-year period, 
we find $R^2_{adj}\approx$ 18\% and 19.5\% for the GARCH(1,1) model
without and with trends, respectively. For longer-term predictions,
one expects the $R$-squared's to grow further as for the EWMA model (subsection 3.6).

\subsection{Models based on Price Ranges}

It has been proposed that the square of the range of intraday prices is a more efficient estimator 
of the unobservable current variance than the square of the close-to-close daily return \cite{parkinson}.
To test whether trends can also improve the predictive power of models based on price ranges,
we have repeated the analysis of the GARCH(1,1) model (\ref{garch11}), but with replacement of
the squared returns $r_t^2$ by the squared ranges $R_t^2$, based on the high-and low-prices $H_t$ and $L_t$ on day $t$:  
\begin{equation}
R_t^2=N\Big({H_t-L_t\over H_t+L_t}\Big)^2. 
\label{range}\end{equation}
Here, the normalization factor $N$ is chosen for each asset such that 
the expectation value of $R^2$ is equal to 1. Thus, the range-based GARCH(1,1) model is
\begin{equation}
R_{t}=\sigma_{t}\epsilon_{t},\ \ \ \sigma_{t+1}^2=\omega+\alpha R_t^2+\beta\sigma_t^2+\lambda(\sigma_t^2-1)\phi_{\text{agg},t}+\zeta\phi_{\text{agg},t}^{(2)},
\label{range11}\end{equation}
in analogy with range-based stochastic volatility models \cite{alizadeh}.
The regression results are shown in table 9 (left columns). 
A comparison with table 8 shows that, as expected, the squared range 
can be forecast with higher $R^2_{adj}$ than the squared return.
Importantly, our results confirm that 
including medium-term trends as explanatory factors can also 
significantly improve the volatility forecasts of the range-based GARCH(1,1) model.\\

\begin{table}[h!]
\centering
\begin{tabular}{ |p{1.2cm}|p{1.3cm}|p{1cm}|p{1.3cm}|p{1cm}||p{1.2cm}|p{1.3cm}|p{1cm}|p{1.3cm}|p{1cm}|}	\hline
	 \multicolumn{3}{|l|}{Range GARCH(1,1)} &\multicolumn{2}{|l||}{With Trends} & \multicolumn{3}{|l|}{Range HAR}& \multicolumn{2}{|l|}{With Trends}\\	\hline
	Coeff.& Value  &t-stat& Value  &t-stat & Coeff. & Value  &t-stat &Value  &t-stat\\	\hline
$\omega$      &$  +0.07$&3.3  &$+0.06$&4.0  &   $c$      &$+0.22$&6.0 &$ +0.08$&3.1   \\
$\alpha$    &$  +0.18$&8.0  &$+0.20$&6.7  &   $\beta_d$&$+0.18$&2.9 &$ +0.17$&3.2   \\
$\beta$      &$  +0.75$&29   &$+0.61$&15   &   $\beta_w$&$+0.28$&4.7 &$ +0.25$&4.3    \\
         &-         &  -   &-       &     &   $\beta_m$&$+0.32$&6.4 &$ +0.22$&5.1    \\
$\lambda$      &-         &-    &$-0.03$&4.1  &   $\lambda$      &-       &-   &$ -0.15$&7.2    \\
$\zeta$     &-         &-    &$+0.16$&3.3  &   $\zeta$     &-       &-   &$ +0.34$&11  \\ \hline\hline
$R^2$    &$  2044$  &  2.9&$ 2144$ & 3.0 &   $R^2$    &$2105$  &3.0 &$  2210$ & 3.0     \\
$R^2_{adj}$&$1940$  &  2.9&$ 2015$ & 2.9 &   $R^2_{adj}$&$2057$&2.5&$  2156$ & 3.6     \\ 
$\Delta R^2_{adj}$&- &  -   &$75 $   & 3.9 &   $\Delta R^2_{adj}$&-&- &$99$ & 5.8     \\  \hline
\end{tabular}
\caption{Medium-term trends can also improve variance forecasts of range-based models
such as the range-based GARCH(1,1) model (left) or the HAR model (right). The last line shows 
the increase of the out-of-sample $R$-squared relative to the models without trends.}
\label{tab:all_var_reg}
\end{table}

Another popular range-based model is the HAR model \cite{corsi}, which we implement
using (\ref{range}), and to which we also add trends
as explanatory factors.
To preserve the simplicity of the model as much as possible, in this case
we use the factors of the simplest model (\ref{volatrend}):
$$ R_{t+1}^2=c+\beta_d R_t^2+\beta_w {1\over5}\sum_{i=0}^4 R_{t-i}^2+\beta_m {1\over22}\sum_{i=0}^{21} R_{t-i}^2+\lambda\phi_{\text{agg},t}+\zeta\phi_{\text{agg},t}^{(2)}+\epsilon_{t+1}^2. $$
Here $R_{t,5}$ and $R_{t,22}$ are the equally-weighted weekly (5-day) and monthly (22-day) averages of the 
range-squared. The regression results with and without trends are also shown in table 9 (right columns).
As observed for the other models, we see that trends can significantly increase the out-of-sample $R^2$.
As in the case of the GARCH(1,1) model, we have verified that the $R^2$'s can be further 
increased by applying the model to sub-sets of our data.
E.g., for equity indices we find out-of-sample $R^2$'s over the whole period close to 50\%.

\subsection{Stochastic Volatility Models}

In stochastic volatility (SV) models \cite{hull, harvey}, the variance $\sigma^2_t$ 
is a new random variable with its own noise, in addition to the asset price. This includes  
the Heston model \cite{heston}:
$$r_t=\sigma_tz_t\ \ \ ,\ \ \ \sigma^2_{t+1}-\sigma^2_t=\kappa(\theta-\sigma^2_t) +\xi\sigma_t\epsilon_t.$$
Here, the two noise terms $z_t$ for the asset price and $\epsilon_t$ for the variance 
are assumed to have have variance 1
and correlation $\rho=$cor$(z_t,\epsilon_t)$, where one expects $\rho<0$ to reflect the leverage effect. 
Let us refine the Heston model by including the aggregate trend factors (\ref{aggregate}):
$$\sigma^2_{t+1}-\sigma^2_t=\kappa(\theta-\sigma^2_t)+\lambda\phi_{\text{agg},t}+\zeta\phi_{\text{agg},t}^{(2)}+\xi\sigma_t\epsilon_t.$$
Since OLS regression cannot be used for such models, 
we fit the model parameters by maximum likelihood estimation and use the out-of-sample quasi likelihood per observation
to measure the predictive power.
The results are shown in table 10. We see that $\lambda$ and $\zeta$ 
are statistically highly significant, partly at the expense of $(1-\kappa)$. 
The leverage effect is now reflected both in the negative correlation $\rho$ and
in the linear trend coefficient $\lambda$ (if one set the correlation $\rho$ to zero, 
$\lambda$ would compensate for it by becoming negative).
The increase of quasi-likelihood per observation by $0.0088$ is statistically highly significant. 
Thus, trends can also improve the predictive power of the Heston model.\\

Instead of the variance, one may alternatively model the log-variance as a new random variable,
as in the Taylor model \cite{taylor}:
$$ r_t=\sigma_t z_t,\ \ \ ,\ \ \ \sigma_t^2=e^{h_t}\ \ \ ,\ \ \ h_{t+1}-\mu=\gamma+\beta (h_t-\mu) + \xi\epsilon_t.$$
In the original Taylor model, the correlation $\rho=$cor$(z_t,\epsilon_t)$ is assumed to be zero;
the Leverage SV model \cite{jacquier, yu} generalizes this to nonzero correlation
in analogy with the Heston model. We have already seen in fig. 4 (right) that, 
in the absence of trends, the next-day variance can possibly be modeled more accurately in 
terms of its logarithm:
$$\langle\ln\sigma_{t+1}^2\rangle\ \approx\ \tilde d\cdot \ln \sigma_t^2\ \ \ 
\text{with}\ \ \ \tilde d\ <\ 1,$$
where $\sigma_t^2$ was the EWMA variance and the brackets denote the expectation value.
The LSV model precisely implements this alternative approach, but with the latent log-variance $h_t$. 
Let us refine the LSV model as well by including trends as factors:
$$h_{t+1}-\mu=\beta(h_t-\mu) +\lambda\phi_{\text{agg},t}+\zeta\phi_{\text{agg},t}^{(2)}.$$
The MLE results are also shown in table 10 (right columns). 
They are quite similar to those for the Heston model: 
both $\lambda$ and $\zeta$ are statistically highly significant.
Thus, trends can also improve the predictive power of the Leverage SV model.\\

\begin{table}[h!]
\centering
\begin{tabular}{ |p{1.7cm}|p{1.5cm}|p{0.8cm}|p{1.3cm}|p{0.8cm}||p{1.7cm}|p{1.5cm}|p{0.8cm}|p{1.3cm}|p{0.8cm}|}	\hline
	\multicolumn{3}{|l|}{Heston model} &\multicolumn{2}{|l||}{With Trends}& \multicolumn{3}{|l|}{LSV model} & \multicolumn{2}{|l|}{With Trends}\\	\hline
	Coeff. & Value  &t-st.& Value  &t-st. &Coeff. & Value  &t-st. &Value  &t-st.\\	\hline
$\theta$ &$+0.938$ &13 &$+0.930$ &16 &$\mu$     &$-0.382$&5.6&$-0.382$& 10\\
$\kappa$ &$+0.018$ &6.0&$+0.042$ &5.3&$\beta$    &$-0.018$&6.0&$-0.040$& 14\\
$\rho$   &$-0.223$ &8.6&$-0.293$ &7.5&$\rho$    &$-0.206$&8.2&$-0.254$& 9.1\\
$\xi$    &$+0.134$ &8.4&$+0.156$ &8.2&$\xi$     &$+0.150$&10 &$+0.176$& 13\\
$\lambda$&-        & - &$+0.006$ &4.3&$\lambda$ &   -    &-  &$+0.007$&4.3\\
$\zeta$  &-        &-  &$+0.025$ &8.9&$\zeta$   &  -     &-  &$+0.025$& 19\\  \hline\hline
$1-Q/N$  &0.2315   &-  & 0.2406  & - &$1-Q/N$   &0.2404  &-  &0.2485  & -\\
Out-of-s.&0.2307   &-  & 0.2395  & - &Out-of-s. &0.2398  &-  &0.2479  &  -\\
$\Delta Q/N$ &-        &-  & 0.0088  &8.4&$\Delta Q/N$  &-       &-  &0.0081  & 12\\  \hline
\end{tabular}
\caption{Trends can also improve volatility forecasts of the Heston- and Leverage Stochastic Volatility (SV) models, based on maximum likelihood estimation. Here, daily returns are normalized to have standard deviation 1, and the coefficients are not annualized. $Q/N$ is QLIKE per observation.
As its level is arbitrary, no $t$-statistics are reported for it. 
The last line shows the improvement of $Q/N$
relative to the models without trends, and its $t$-statistics.}
\label{tab:all_var_reg}
\end{table}

We leave it for future work to examine whether trends can also add value to 
other stochastic volatility models such as the multi-fractal random walk \cite{bacry, borland}.

\section{Summary and Conclusions}

Let us summarize our results.
In analogy with our previous model (\ref{cubic})
of expected next-day returns $r_{t+1}$ in financial markets, 
we have modeled the next-day variance $\sigma_{t+1}^2$ based on the current day's trends.
While tomorrow's expected return is modeled by a cubic polynomial
in today's trend strength $\phi_t$, second-order polynomials in $\phi_t$ seem to be sufficient to model tomorrow's variance.
More precisely, we use the linear and quadratic trend strengths $\phi_{\text{agg}}$ and
$\phi_{\text{agg}}^{(2)}$, aggregated across trend horizons (\ref{aggregate}).
However, today's variance $\sigma_t^2$ must also be included as an
explanatory factor, as market variances come in clusters and tend to revert to their longer-term means. 
In fact, $\sigma^2_t$ turns out to be the dominant factor.
The question then becomes to what extent its predictive power can be enhanced by the trend factors.\\

We have first addressed this question 
at the example of a simple linear EWMA-based variance model, which leads to 
the following forecast of the square of the next-day return:
\begin{eqnarray}
r_{t+1}^2&=&a+d\cdot \sigma_t^2+e\cdot \phi_{\text{agg},t}+f\cdot \phi_{\text{agg},t}^{(2)}+\epsilon_{t+1} \ \ \notag\\ 
\text{with}&&a=0.05\pm0.03,\ \ \ d = 0.67\pm0.07,\ \ \ e = -0.21\pm0.03,\ \ \ f = 0.34\pm0.04.\notag
\end{eqnarray}
The $a$ and $d$ terms are already present in ordinary linear EWMA-based variance 
models, where $d-1$ is negative and measures how quickly the variance reverts to its equilibrium value,
which in our case has been normalized to 1. \\

The $e$ and $f$ terms refine such models by also taking current trends into account. 
The positive value of $f$ implies that the variance tends to grow day after day in 
times of strong trends. This explains why it is high after trends have built up (fig. 1, right).
The negative value of $e$ shows that the variance grows faster in times of strong down-trends
as opposed to up-trends. Since trends measure cumulative recent returns, this can be interpreted 
in terms of the ``leverage effect": strong negative returns tend to be followed by an increase of the variance.
We have confirmed that this effect is particularly strong for equities and mostly stems
from trends with trend horizons of up to several weeks.  \\

Adding the linear and quadratic trend factors to the model yields a modest, 
but statistically highly significant improvement of the out-of-sample $R$-squared in the case of 
ordinary least squares regression, and of the quasi-likelihood per observation in the case of
maximum likelihood estimation. This improvement is higher, and thus economically
more significant, for forecasting horizons that are longer than 1 day (up to 3 months).
Within their estimation errors, our results are stable over the 35-year time period of our analysis.\\

We have also carried out a less detailed analysis for advanced variance models:
the GARCH(1,1) model, the GJR GARCH model, the range-based GARCH(1,1) model, the HAR model,
the Heston model and the Leverage Stochastic Volatility model.
In each case, the conclusion is the same as for the EWMA model:
adding the current trend strengths and their squares on multiple time horizons 
as explanatory factors for the next-day variance yields a moderate but statistically 
highly significant improvement of the predictive power.\\

Let us propose a general explanation for this improvement: all of these advanced models include
today's squared return as a key explanatory factor for tomorrow's variance.
However, as is well-known among practitioners, financial markets are characterized 
by approximate scale invariance.
In particular, market trends are empirically observed on all scales from a few days to several months,
and the persistence $b$ of trends and the strength of reversion $c$ in (\ref{cubic}) 
are remarkably similar across all trend horizons. 
In view of this approximate scale invariance,
one indeed expects that the recent $T$-day return and its square 
should enter a variance forecast in a similar way as the recent 1-day return.
This recent $T$-day return is just what is measured by the $T$-day trend.\\

The fact that current trends can help to predict future returns 
has successfully been exploited for decades by the trend-following industry. 
Our observation that trends can also help to predict future
variances should allow for a refinement of market risk management tools. 
However, our results might also have a wider significance.
Each of the different variance models we have tested here
contains its own set of parameters that must be fitted to 
historical data, which brings with it the risk of overfitting the data. 
It would of course be very helpful to have a fundamental model of financial markets 
that can not only replicate the empirically observed 
stylized facts of finance \cite{mant, cont, sara2},
but also derive the true stochastic processes for returns, variances 
and correlations from first principles.\\

In this context, let us conclude on a speculative note.
The previous results (\ref{cubic})
have led to the proposal of a lattice gas model of financial markets \cite{me2}. 
The lattice represents the social network of investors, 
while the gas molecules represent the shares of an asset (such as a stock) 
that are distributed across this network.
In efficient markets \cite{fama}, arbitrageurs should drive this gas to its critical temperature,
where it is indeed characterized by approximate scale invariance.
This model can replicate the observed interplay (\ref{cubic}) of trends and trend reversals. 
It will be very interesting to work out the precise stochastic process that it predicts
for the variance, and to compare with the empirical observations reported here. 

\section*{Acknowledgements} 

We would like to thank Ashkan Nikeghbali, 
Gisela Reichmuth, and Stefan Ask for discussions.
This research is supported by the Swiss National Science Foundation under 
grant no. PT00P2\_206333. \\

The authors used ChatGPT (OpenAI) to assist with reviewing the literature on volatility models, 
and to find errors in the R code, which was independently written by the authors. Claude (Anthropic) was used for the computationally intensive fitting of the parameters of the
stochastic volatility models in subsection 4.3. 
All analyses were performed by the authors, who take full responsibility for the content of the published article.

\end{document}